\documentclass[aps,prl,twocolumn,showpacs]{revtex4}

\usepackage{here}
\usepackage{graphicx}

\newcommand\pictc[5]{\begin{figure}
                       \centerline{
\includegraphics[width=#1\columnwidth,keepaspectratio,height=0.8\textheight]{#3}}
                   \protect\caption{\protect\label{fig:#4} #5}
                    \end{figure}            }
\newcommand\pict[4][1.]{\pictc{#1}{!tb}{#2}{#3}{#4}}
\newcommand\rpict[1]{\ref{fig:#1}}

\newcommand\leqt[1]{\protect\label{eq:#1}}
\newcommand\reqtn[1]{\ref{eq:#1}}
\newcommand\reqt[1]{(\reqtn{#1})}

\begin{document}
\begin{sloppy}

\title{Nonlinear Bloch-wave interaction and Bragg scattering in optically-induced lattices}

\author{Andrey A. Sukhorukov$^1$, Dragomir Neshev$^1$, Wieslaw Krolikowski$^2$, and Yuri S. Kivshar$^1$}

\affiliation{$^1$Nonlinear Physics Group and $^2$Laser Physics
Center, Center for Ultra-high bandwidth Devices for Optical
Systems (CUDOS), Research School of Physical Sciences and
Engineering, Australian National University, Canberra ACT 0200,
Australia} \homepage{http://www.rsphysse.anu.edu.au/nonlinear}

\begin{abstract}
We study, both theoretically and experimentally, the Bragg scattering of light in optically-induced photonic lattices and 
reveal the key physical mechanisms which govern nonlinear self-action of narrow beams under the combined effects of Bragg scattering and wave
diffraction, allowing for selecting bands with different effective dispersion.
\end{abstract}


\maketitle

{\em Dispersion} and {\em diffraction} are the fundamental
phenomena of wave physics, and they are responsible for temporal
pulse spreading and beam broadening. The rate of wave broadening
can be controlled in periodic structures where the effective
geometric dispersion provides a key physical mechanism for
manipulating waves in various physical systems, including the
Bragg gratings in optical
fibers~\cite{Agrawal:1989:NonlinearFiber}, waveguide
arrays~\cite{Eisenberg:2000-1863:PRL}, and Bose-Einstein
condensates (BECs) in optical
lattices~\cite{Eiermann:2003-60402:PRL}.

A strong change of the wave dispersion and diffraction occurs in
the vicinity of spectrum band gaps where waves experience resonant
Bragg scattering from a periodic
structure~\cite{Yeh:1988:OpticalWaves}. In the structure shown in
Fig.~\rpict{scatter}, both scattering and spreading depend on the
angle between the beam and grating. For broad beams,
the evolution of transmitted and scattered waves is defined by the
effective dispersion parameters at the spectral
peaks~\cite{Sipe:1988-132:OL}, and can be studied in the
effective-mass approximation~\cite{Ashcroft:1984:SolidState}.
The dynamics of narrow beams and wavepackets becomes more complicated, and it is strongly affected by nonlinearity, as recently discussed in Ref.~\cite{Eiermann:2003-60402:PRL}.
However, the key physical mechanisms which govern nonlinear self-action of narrow beams under {\em combined effects of the Bragg scattering and wave diffraction} remain largely unexplored.

In this Letter we study, both theoretically and experimentally,
light propagation in optically-induced photonic lattices near the
Bragg scattering angle. For the first time to our knowledge, we
reveal generic relations between the Bloch-wave spectrum of the
periodically modulated media and the specific structure of the
diffracted waves describing analytically the key patterns of the
beam scattering and self-action observed in our experiments. We
also study the nonlinear Bloch-wave interaction and observe
self-focusing in the spectrum gaps tracing the signatures of the
Floquet-Bloch solitons.

\pict[0.6]{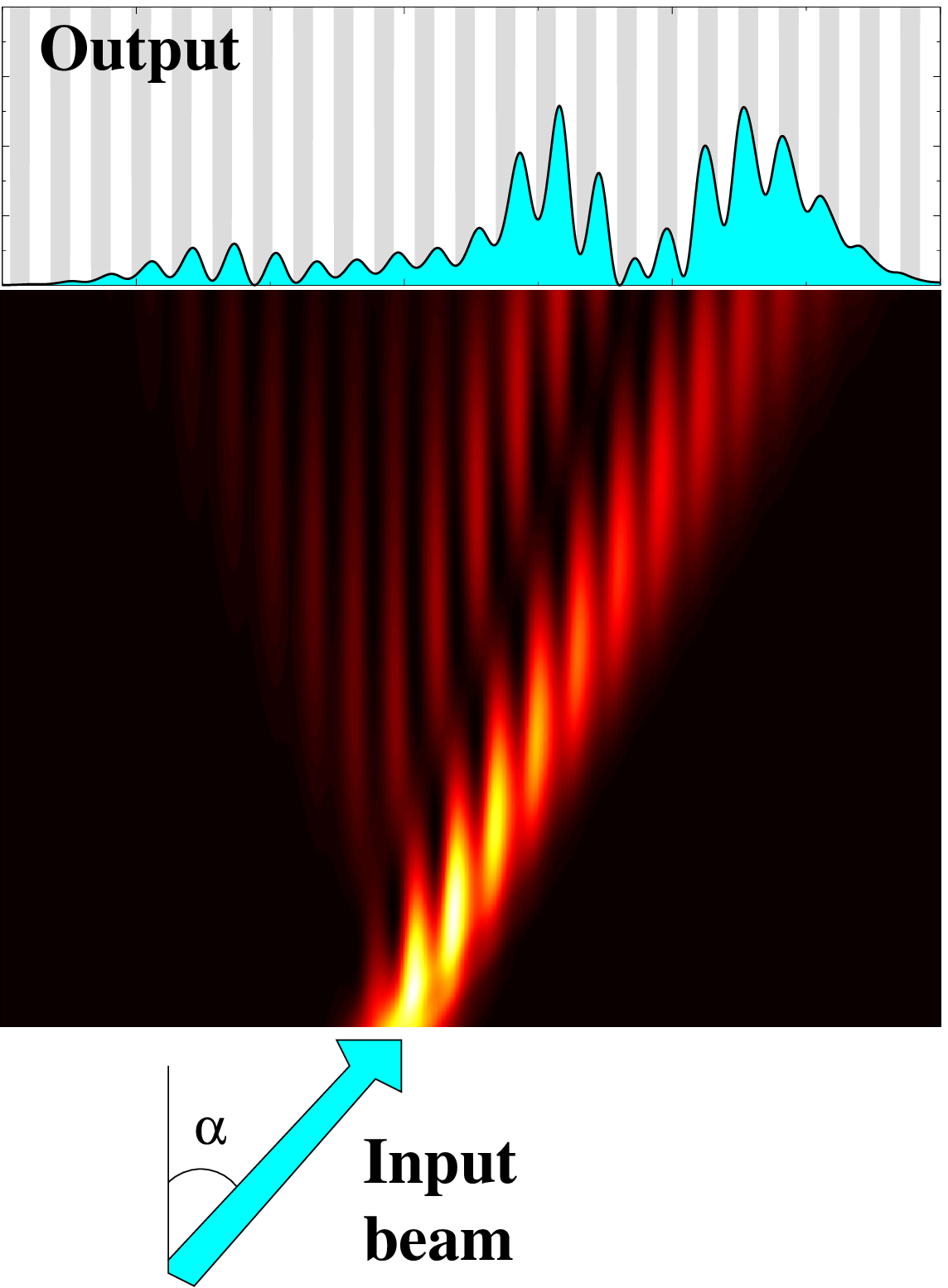}{scatter}{Schematic of a laser beam
propagating in a periodic photonic structure. The input Gaussian
beam is split into several Bloch modes due to the Bragg scattering; top plot shows the output profile superimposed on top of the grating.
Beam evolution is defined by the inclination angle $\alpha$.}

{\em Model.} Essential features of light scattering in periodic
photonic structures and the BEC dynamics in optical lattices can
be described by similar mathematical models. For definiteness, we
consider the optical beam propagation in a one-dimensional
periodic structure described by the dimensionless equation,
\begin{equation} \leqt{NLS}
   i \frac{\partial E}{\partial z}
   + D \frac{\partial^2 E}{\partial x^2}
   + \nu(x) E
   + {\cal F}( x, |E|^2) E
   = 0 .
\end{equation}
where $E(x,z)$ is the normalized electric field envelope, $x$ and
$z$ are the transverse and propagation coordinates, normalized to
the characteristic values $x_s$ and $z_s$, respectively, $D = z_0
\lambda / (4 \pi n_0 x_0^2)$ is the beam diffraction coefficient,
$n_0$ is the average medium refractive index, and $\lambda$ is the
vacuum wavelength. Function $\nu(x)$ characterizes the periodic
modulation of the optical refractive index,
$\nu(x+d)\equiv\nu(x)$, where $d$ is the spatial lattice period.
The function ${\cal F}( x, |E|^2)$ defines the self-induced
changes of the refractive index.

The theoretical analysis presented below is equally applicable to
the BEC dynamics including the recent experimental results for the
BEC dispersion management in optical
lattices~\cite{Eiermann:2003-60402:PRL}. In this case, $E(x,z)$ stands for the mean-field condensate wave function, $x$ is the spatial direction along a cigar-shaped trap, $z$ is the normalized time, and $D$ is a constant parameter. The periodic function $\nu(x)$ defines the effective potential of an optical lattice, and ${\cal F}( x, |E|^2) \simeq \gamma |E|^2$ characterizes the inter-atomic scattering.

{\em General analytical approach.} First, we present our general
theory for describing scattering and diffraction of optical beams
in periodic photonic structures [see Fig.~\rpict{scatter}] based
on the Floquet-Bloch formalism~\cite{Yeh:1988:OpticalWaves}. The
Bloch waves are special solutions of the linearized
equation~\reqt{NLS} of the form, $E_{\kappa,n}(x,z) =
\psi_{\kappa,n}(x) \exp(i \kappa x + i \beta_{\kappa,n} z)$, where
$\beta_{\kappa,n}$ and $\kappa$ are the Bloch-wave propagation
constant and wavenumber, respectively, and the index $n =
1,2,\ldots$ marks the transmission band. The Bloch wave
$\psi_{\kappa,n}(x)$ has the periodicity of the photonic
structure, $\psi_{\kappa,n}(x+d) \equiv \psi_{\kappa,n}(x)$, and
this constraint defines a specific dispersion relation
$\beta_{\kappa,n}$, as sketched in Fig.~\rpict{blochSpectr}(a). In
a one-dimensional structure, the transmission bands are separated
by gaps~\cite{Yeh:1988:OpticalWaves}, where the wave propagation
is forbidden ($\kappa$ is real). The band-gaps are associated with
Bragg reflection of light at certain incident angles.

\pict{fig02.eps}{blochSpectr}{(a)~Dispersion curves for the
Bloch waves in the first and second transmission bands. Shading
marks the band gaps. (b,c)~Propagation angle and diffraction
coefficients of the Bloch waves. (d)~Excitation coefficients of
the Bloch waves; shading shows the normalized Fourier spectrum
($|F(k)|^2$) of a beam incident at the Bragg angle. All parameters
correspond to the experimental conditions.}

The Bloch waves form a complete orthogonal set of eigenfunctions,
and the linear evolution of an input beam in a periodic lattice is
fully defined by its {\em Bloch-wave spectrum} $B(\kappa,n)$.
Indeed, the electric field envelope $E$ can be found at a fixed
$z$ as follows
\begin{equation} \leqt{field}
   E(x,z) = \sum_n \int_{-\pi/d}^{+\pi/d}
               B(\kappa,n) \psi_{\kappa,n}(x)
                  e^{i \kappa x + i \beta_{\kappa,n} z} d\kappa.
\end{equation}
The Bloch-wave spectrum can be linked to the Fourier spectrum of
the input beam,
\begin{equation} \leqt{ispectr}
   B(\kappa,n) = \sum_m C_n(\kappa+2 \pi m /d)
    F(\kappa+2 \pi m /d) ,
\end{equation}
where $F(k) = (2\pi)^{-1} \int_{-\infty}^{+\infty} E(x,0) \exp(-i
k x) dx$, and $C_n(\kappa)$ are the Bloch-wave excitation
coefficients, as shown in Fig.~\rpict{blochSpectr}(d).

In order to capture the essential features of the beam diffraction
in a periodic lattice, we calculate the beam profiles in the far
field regime. Assuming that $z$ exceeds several diffraction lengths, we use the stationary-phase approximation~\cite{Born:2002:PrinciplesOptics} to separate the leading-order contributions in the integral representation~\reqt{field} and obtain
\begin{equation} \leqt{fieldSpectr}
   E(x,z) \simeq \sqrt{\pi} \sum_{\kappa, n: x=-z \beta^{(1)}_{\kappa,n}}
               \frac{B(\kappa,n) \psi_{\kappa,n}(x)}{(|\beta^{(2)}_{\kappa,n}| z)^{1/2}}
                e^{ i \varphi} ,
\end{equation}
where $\varphi = \kappa x + \beta_{\kappa,n} z
               + (\pi/4) {\rm sign}(\beta^{(2)}_{\kappa,n})$, and
$\beta^{(j)}_{\kappa,n} =\partial^j \beta_{\kappa,n}/\partial
\kappa^j$. Solution~\reqt{fieldSpectr} allows us to link the spatial
profile of the diffracted beam and its Bloch-wave spectrum, and it
provides a nontrivial generalization of the well-known fact that
in a homogeneous medium the profile of a diffracted beam maps
directly its Fourier spectrum~\cite{Born:2002:PrinciplesOptics}.
According to the summation condition in Eq.~\reqt{fieldSpectr},
$\beta^{(1)}_{\kappa,n}$ defines {\em the propagation angle} (or
the group velocity) of the Bloch waves. The second derivative of
the Bloch-wave dispersion, $\beta^{(2)}_{\kappa,n}$, characterizes
the rate of the beam spatial divergence, and it has a meaning of
the {\em effective diffraction coefficient}. In the regions where
$\beta^{(2)}_{\kappa,n} < 0$, the beam diffraction is {\em normal}
(as in a homogeneous medium), and it is {\em anomalous} for
$\beta^{(2)}_{\kappa,n} > 0$. As illustrated in
Figs.~\rpict{blochSpectr}(a-c), the group velocity vanishes at the
band edges, and it reaches a maximum absolute value inside the
band, whereas the beam diffraction changes from normal to
anomalous.  A fundamental conclusion of our analysis is the
following: at any given spatial location $x$, the beam profile is
found as {\em a superposition of Bloch waves from different
spectral bands}, and each of the bands gives {\em two
contributions} associated with both {\em normal} and {\em
anomalous} diffraction. Nonlinearity produces the opposite effects
on such waves, resulting in either self-focusing or
self-defocusing, and it is the competition of these effects that
will define the beam shaping in periodic structures.

Equation~\reqt{fieldSpectr} becomes formally invalid in the
vicinity of a zero-diffraction point, which corresponds to {\em a
diffraction catastrophe}. The corresponding contribution to the
total field can be expressed through the Airy functions by taking
into account higher-order diffraction terms. When
$\beta^{(2)}_{\kappa,n} = 0$, the general expression can be
simplified, $- 2 \pi z^{-1/3} B(\kappa,n) \psi_{\kappa,n}(x)
|6/\beta^{(3)}_{\kappa,n}|^{1/3} [\Gamma(-1/3)]^{-1}$. The
contribution from the zero-diffraction region may become dominant
in the far field where $B(\kappa_1,n) |z
\beta^{(3)}_{\kappa_1,n}|^{1/3} \ll B(\kappa_2,n) |z
\beta^{(2)}_{\kappa_2,n}|^{1/2}$. This effect is responsible for
the formation of a specific diffraction pattern with two side
lobes in waveguide arrays~\cite{Eisenberg:1998-3383:PRL}. In the
case of dynamically induced
gratings~\cite{Fleischer:2003-23902:PRL, Neshev:2003-710:OL}, the
refractive index contrast is smaller, and the spectral regions
with near-zero diffraction are relatively narrow [see
Fig.~\rpict{blochSpectr}(c)] and are not expected to make a
dominant contribution for experimentally feasible propagation
lengths.

{\em Experiment vs. theory. } In order to study the Bragg
scattering and Bloch-wave interaction experimentally, we create an
optically-induced periodic lattice by interfering two
ordinary-polarized coherent laser beams in a biased
photorefractive crystal SBN:60, as discussed in
Refs.~\cite{Fleischer:2003-23902:PRL,Neshev:2003-710:OL}. We apply
an external electric field (up to 4000 V/cm) to the crystal,
creating a periodic modulation of the optical refractive index
through the electro-optic effect. In the SBN crystal, orthogonally
polarized waves have significantly different electro-optic
coefficients. We use the ordinary polarized interfering beams
which propagate linearly and create a stable periodic lattice,
whereas the extraordinary polarized probe beam experiences strong
nonlinear self-action, which can be quantified through the
nonlinear change of the refractive index, $\nu(x) + {\cal F}( x,
|E|^2) = - \gamma (I_b + I_p(x) + |E|^2)^{-1}$, where $I_b$ is the
constant dark irradiance, $I_p(x) = I_g \cos^2(\pi x / d)$ is the
two-beam interference pattern which induces the periodic lattice,
and $\gamma$ is the nonlinear coefficient which is proportional to
the applied DC field. In order to match the conditions of our
experimental observations in the theory and numerical
calculations, we use the following parameters: $\lambda = 0.532
\mu$m, $n_0 = 2.4$, $x_0 = 1\mu$m, $z_0 = 1$mm, $d = 15.1$, $I_b =
1$, $I_g = 0.4$, $\gamma = 2.13$, the crystal length is $L =
15$mm, and the input beam width is $21.16 \mu$m. A detailed
description of our experimental setup can be found in our earlier
paper~\cite{Neshev:2003-710:OL}.

\pict{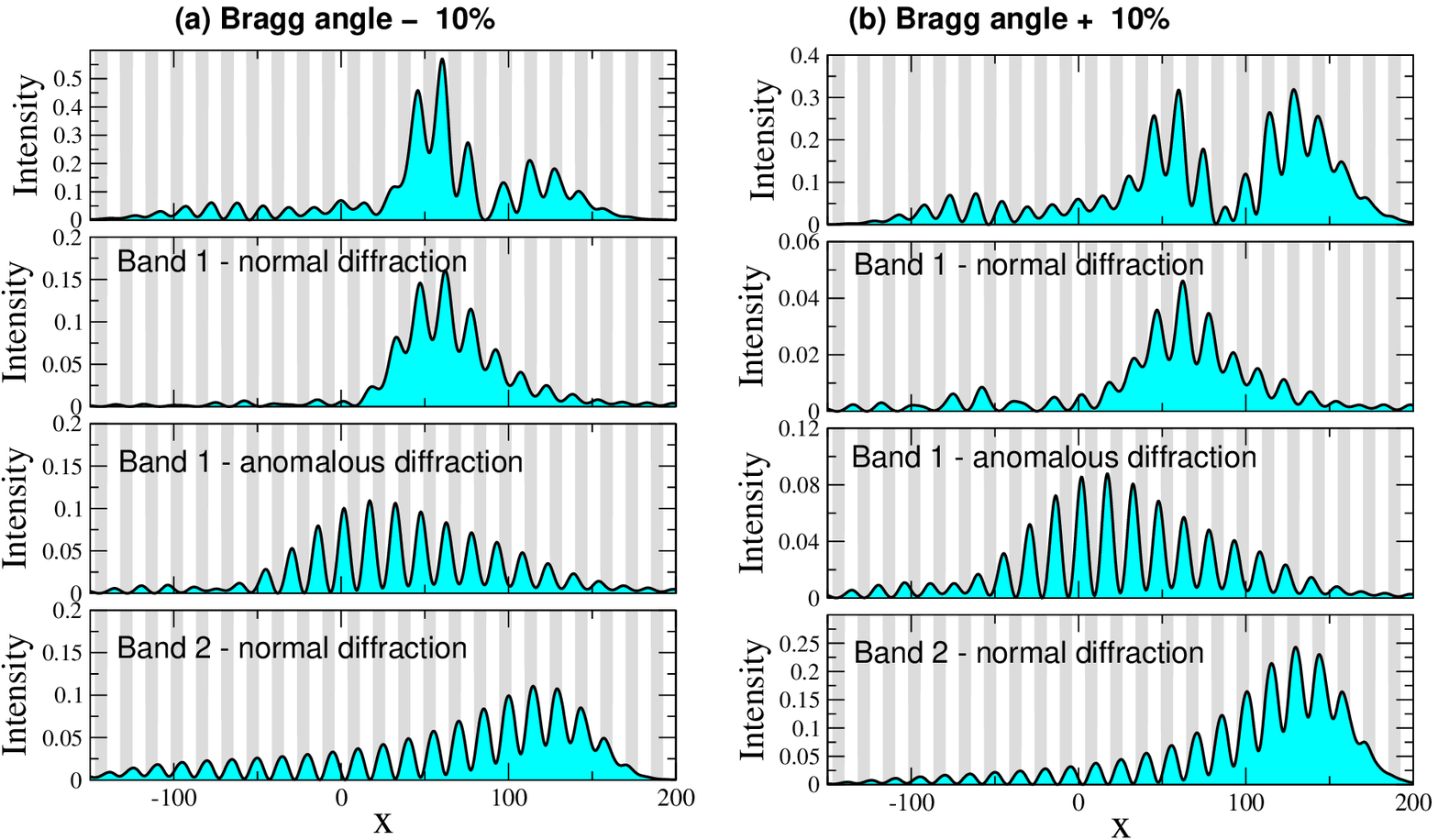}{profSpectr}{ 
Numerically calculated output beam
profiles in the linear regime for an input Gaussian beam incident
at the angles (a) 10\% below and (b) 10\% above the Bragg
resonance. Intensity levels are normalized to the input peak
intensity $I_0$. Shown is decomposition of the total intensity
profile (top) between spectral regions with normal and anomalous
diffraction of Bloch waves in the first and second bands (as
indicated by labels). Shading marks the positions of the grating
minima.}

\pict{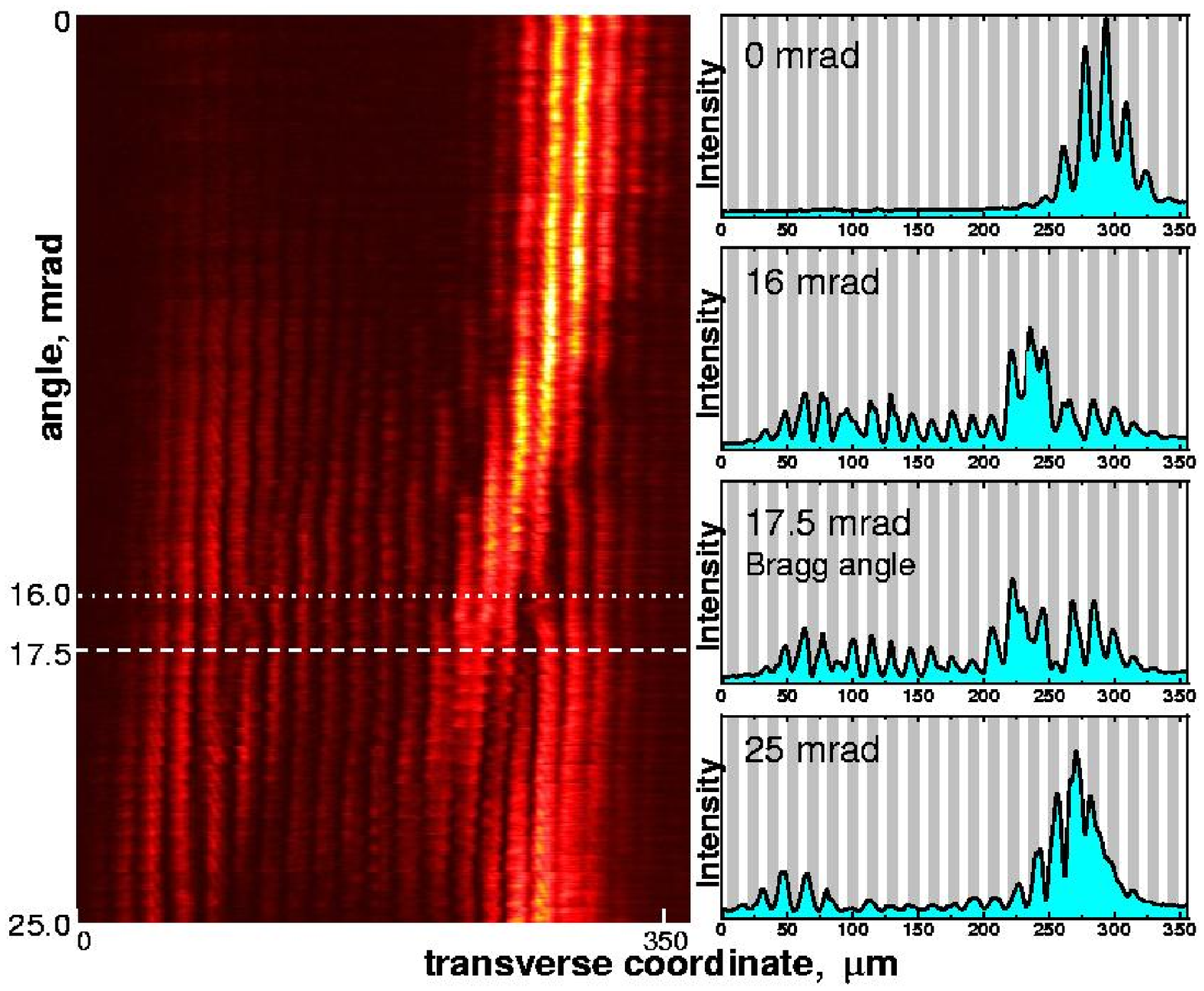}{scan}{ Experimentally observed beam scattering
by an optically-induced lattice for low laser powers. Left: output
intensity vs. the tilt angle; right: beam profiles at different
angles, shading marks the positions of the grating minima.}

First, we perform numerical analysis of beam scattering in the linear regime using the exact Bloch-wave decomposition [Eq.~\reqt{fieldSpectr}]. When the inclination angle of the scattered beam approaches that
of the Bragg resonance, we observe {\em the simultaneous
excitation of several spectral bands} by an input Gaussian beam, as illustrated in Figs.~\rpict{profSpectr}(a,b). These results are in perfect agreement with our general conclusion that a spatial profile of the
diffracted beam maps various Bloch-wave spectral components. Indeed, 
three spectral regions can be excited simultaneously according to Fig.~\rpict{blochSpectr}(d), where we superimpose the Fourier spectrum of the input beam on top of the Bloch-wave excitation coefficients. 

In experiment, first we study the linear scattering when the beam
power is low. Experimental results are summarized in
Fig.~\rpict{scan}, where we show the dependence of the output beam
profile on the tilt angle of the periodic lattice. Close to normal
incidence, the first band is dominant, whereas at larger angles
the excitation of the second band grows. In the vicinity of the
Bragg angle [marked as a dashed line in Fig.~\rpict{scan}(left)],
the transmitted beam consists of two parts which correspond mainly
to the excitation of the first and second bands, as predicted
theoretically (cf. Fig.~\rpict{profSpectr}). For angles larger
than the Bragg angle, the transmitted beam is composed mostly of
the Bloch waves from the second spectral band. We note that the
result of scattering very weakly depends on the initial beam
position. This happens because the spectral width of the input
beam is less than the width of the first Brillouin zone ($\Delta k
< 2 \pi / d$), and according to Eq.~\reqt{ispectr} the beam shift
introduces a uniform phase modulation of the Bloch-wave spectrum
but does not change its shape.

\pict{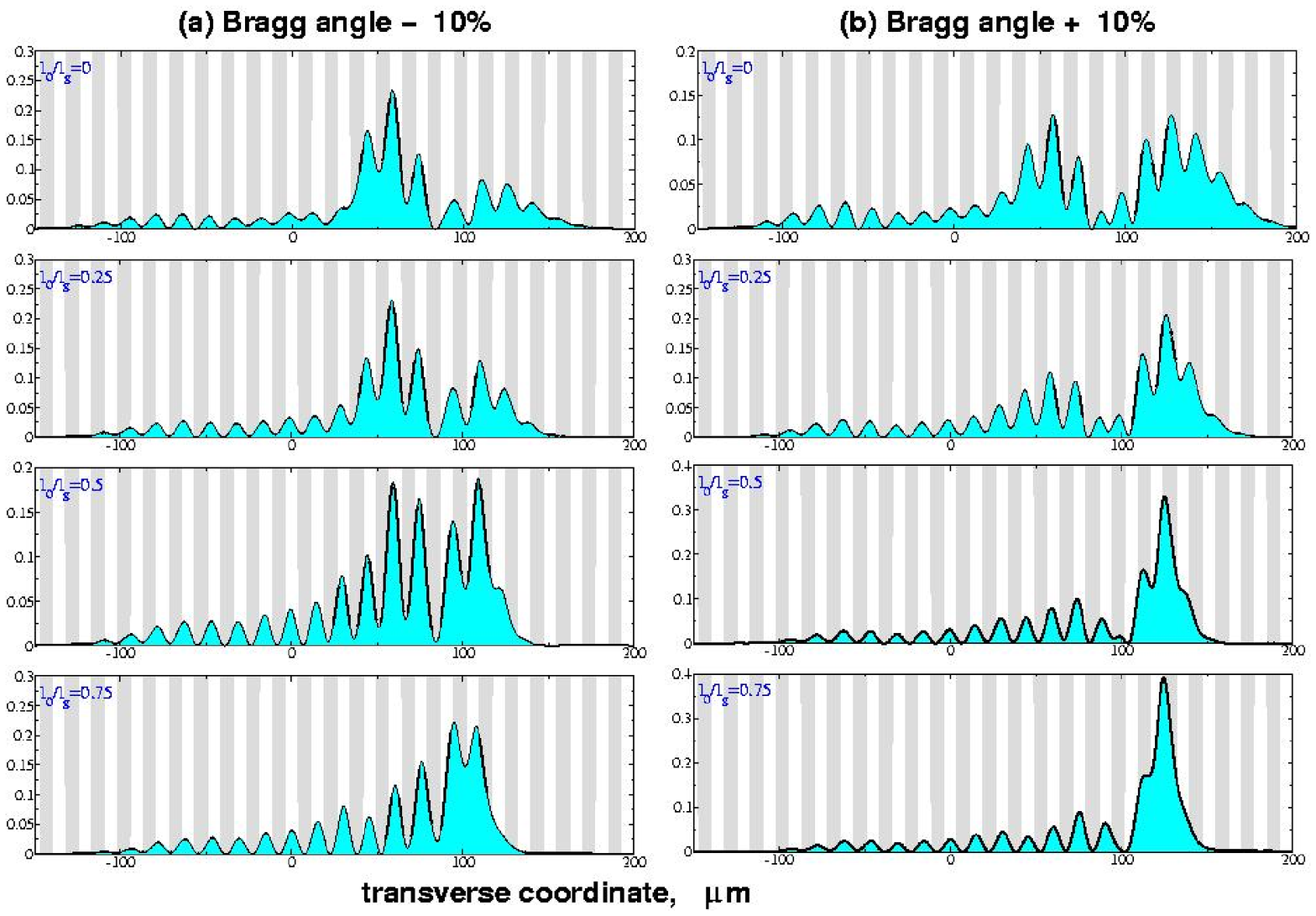}{profNonl}{Numerically calculated output beam
profiles for different values of $I_0$. Beam parameters correspond
to Fig.~\rpict{profSpectr}. }

\pict{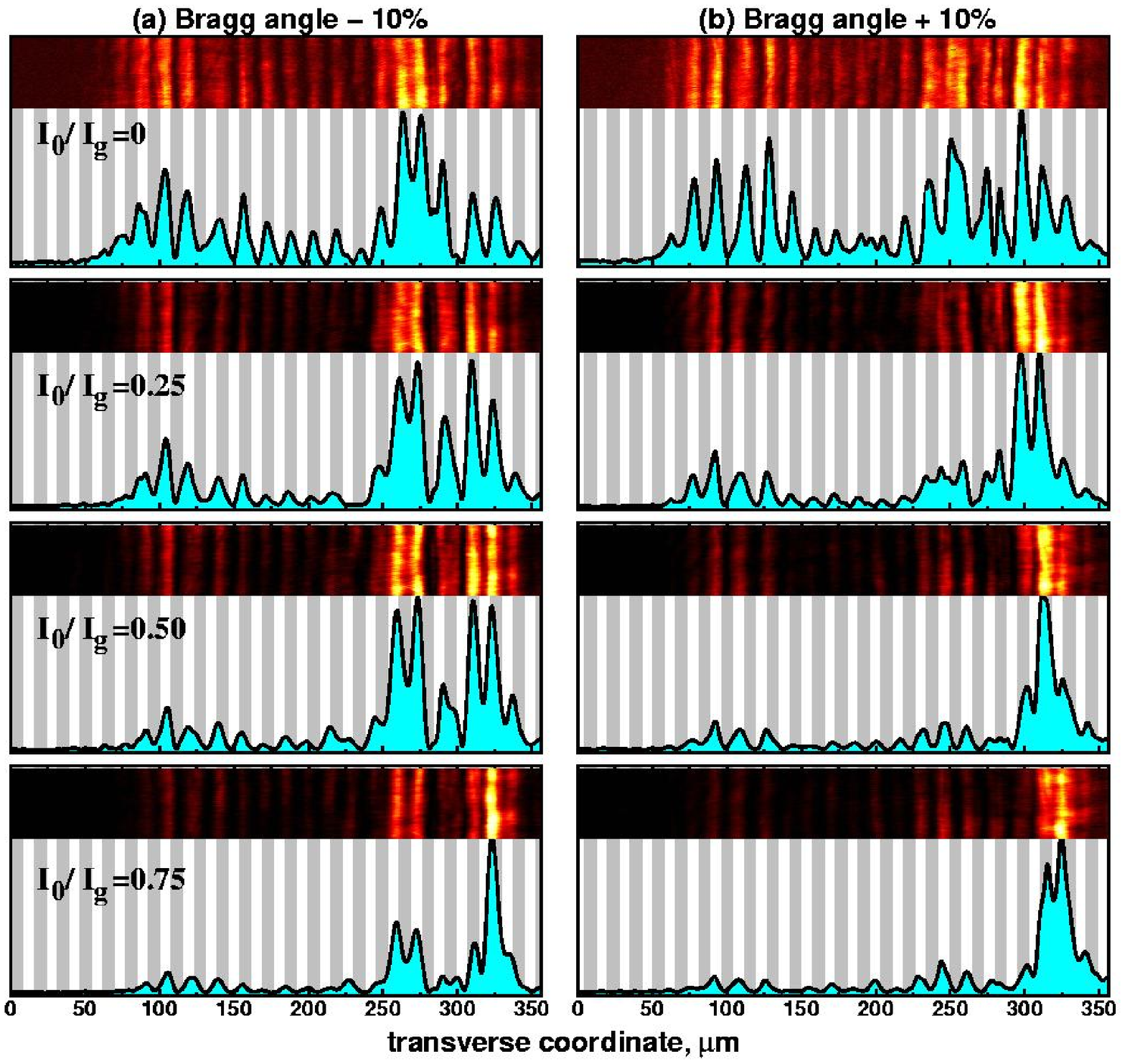}{selfFocus}{Experimentally observed output beam
profiles for different intensities, (a)~below and (b)~above the
Bragg angle.}

As the beam intensity grows, nonlinearity leads to strong changes
in the output beam profiles, see results of numerical simulations of the full model Eq.~\reqt{NLS} in Figs.~\rpict{profNonl}(a,b). In
the nonlinear propagation regime, normally diffracting Bloch-wave
components tend to self-focus whereas anomalously diffracting
components experience self-defocusing. This fundamental property
of the beam scattering suggests a possibility of performing {\em
nonlinear Bloch-wave spectroscopy} of the optically-induced
photonic lattices. In particular, we find that the beam
self-action effect depends strongly on the incident angle. When
the angle is {\em below the Bragg resonance} [see
Figs.~\rpict{profNonl}(a)], the normally diffracting mode is
dominant in the first band, see Fig.~\rpict{profSpectr}(a).
Therefore, as the beam power grows, the mode experiences
self-focusing. At the same time, we observe a strong
nonlinearity-induced energy transfer into the right part of the
beam, which corresponds to the excitation of the second band. In
contrast, for the scattering angles {\em above the Bragg
resonance} [see Figs.~\rpict{profNonl}(b)], the excitation of the
second band is dominant, and in the first band
anomalously-diffracting mode is stronger, see
Fig.~\rpict{profSpectr}(b). This leads to defocusing on the left
part of the transmitted beam, whereas the right part self-focuses.
At higher powers, the beam starts erasing the optically-induced
grating, and scattering is significantly reduced for both incident
angles. In experiment, we increase the intensity of the probe
beam, and observe the transformations of the beam profiles at the
crystal output [see Fig.~\rpict{selfFocus}] in {\em a remarkable
agreement with the theoretical predictions} [cf.
Fig.~\rpict{profNonl}]. Numerical calculations indicate that
either one or two Floquet-Bloch
solitons~\cite{Mandelik:2003-53902:PRL} can form under appropriate
conditions, however the soliton formation distance exceeds the
crystal length due to strong Bloch-wave interactions.

In conclusion, we have studied the Bragg scattering and nonlinear
Bloch-wave inter-band interaction in optically-induced photonic
lattices. We have established generic relations between the
Bloch-wave spectrum and the structure of diffracted waves
describing theoretically the specific patterns of the beam
scattering and self-action observed in experiment.

We thank B. Eggleton, E.~A. Ostrovskaya, and C.~M. de Sterke for useful
discussions. This work was supported by the Australian Research
Council.

\end{sloppy}
\end{document}